\documentclass[aps,preprint]{revtex4}
\usepackage[dvips]{graphics,graphicx}
\usepackage{amsmath}

\begin{document}
\title{Wannier-Stark states in double-periodic lattices II:  two-dimensional lattices}
\author{Evgeny~N.~Bulgakov$^{1}$, Dmitri.~N.~Maksimov$^{1}$, and Andrey~R.~Kolovsky$^{1,2}$}
\affiliation{$^1$Kirensky Institute of Physics, 660036 Krasnoyarsk, Russia}
\affiliation{$^2$Siberian Federal University, 660041 Krasnoyarsk, Russia}
\date{\today}

\begin{abstract}
We analyze the Wannier-Stark spectrum of a quantum particle in tilted two-dimensional lattices with the Bloch spectrum consisting of two subbands, which could be either separated by a finite gap or connected at the Dirac points. For rational orientations of the static field given by an arbitrary superposition of the translation vectors the spectrum is a ladder of energy bands. We obtain asymptotic expressions for the energy bands in the limit of large and weak static fields and study them numerically for intermediate field strength. We show that the structure of energy bands determines the rate of spreading of a localized wave packets which is the quantity measured in laboratory experiments. It is shown that wave-packet dispersion becomes a fractal function of the field orientation in the long-time regime of ballistic spreading.
\end{abstract}

\maketitle

\section{Introduction}
\label{sec1}

In our preceding work \cite{preprint} we have analyzed the Wannier-Stark states (WS-states) in a tilted one-dimensional double-periodic lattice. It was shown that the spectrum of WS-states consists of two Wannier-Stark ladders that affect each other in a rather complicated way. In this paper we extend the analysis of Ref.~\cite{preprint} to double-periodic two-dimensional lattices or, more exactly, to two-dimensional lattices with two sublattices.

The fundamental difference of 2D lattices as compared with 1D lattices is that the Wannier-Stark spectrum and WS-states depend not only on the strength of a static field $F$ but also on its orientation relative to the primary axes of the lattice, where one should distinguish between `rational' and `irrational' orientations. The former are given by an arbitrary superposition of the translation vectors and the latter comprise the remaining directions. For rational orientations WS-states are Bloch-like states in the direction orthogonal to the vector ${\bf F}$ \cite{Naka93,51,92}. Thus they can labeled by the ladder index $n$ and the transverse quasimomentum $\kappa$. Correspondently, the energy spectrum of the system is a ladder of energy bands which are termed Wannier-Stark bands (WS-bands) in what follows. Unlike this situation, for irrational orientations the spectrum is pure point and WS-states are localized states in any direction \cite{92}.

In the work we pay special attention to physical manifestations of the structure of the energy spectrum that can be detected in a laboratory experiment, where two most favorite candidates are wave-guide arrays \cite{Corr13} and cold atoms in optical lattices \cite{Tarr12}. In particular, continuous spectrum of the WS-states for rational orientations of the static field implies ballistic spreading of a localized wave packet. This effect can be easily observed for the light in a wave-guide array and, with some efforts aimed on improving resolution of the imaging system, also for cold atoms in an optical lattice. We analyze the rate of wave-packet  spreading as the function of the static field magnitude and its orientation. The other quantity that can be measured in experiments with cold atoms is the total occupation of the Bloch subbands which change in time because of the interband Landau-Zener tunneling (LZ-tunneling). We shall show that knowledge of  the Wannier-Stark spectrum suffices to predict different regimes of the interband dynamics.   

The structure of the paper is as follows. In Sec.~\ref{sec2} we introduce the model -- the tight-binding Hamilltonian of a square 2D lattice with two sublattices -- and perform preliminary numerical analysis of the Wannier-Stark spectrum. Next, in Sec.~\ref{sec3} we obtain asymptotic expressions for WS-bands in the limits of strong and weak static field.  These results are used in the subsequent Sec.~\ref{sec4}, which compares the rate of ballistic spreading calculated on the basis of WS-bands with that observed in numerical simulations of the wave-packet dynamics, and in Sec.~\ref{sec5}, which discusses Landau-Zener transitions between the Bloch subbands. The main results of the work are summarized in the concluding Sec.~\ref{sec6}.

\section{Wannier-Stark energy bands}
\label{sec2}

\begin{figure}
\center
\includegraphics[width=9cm,clip]{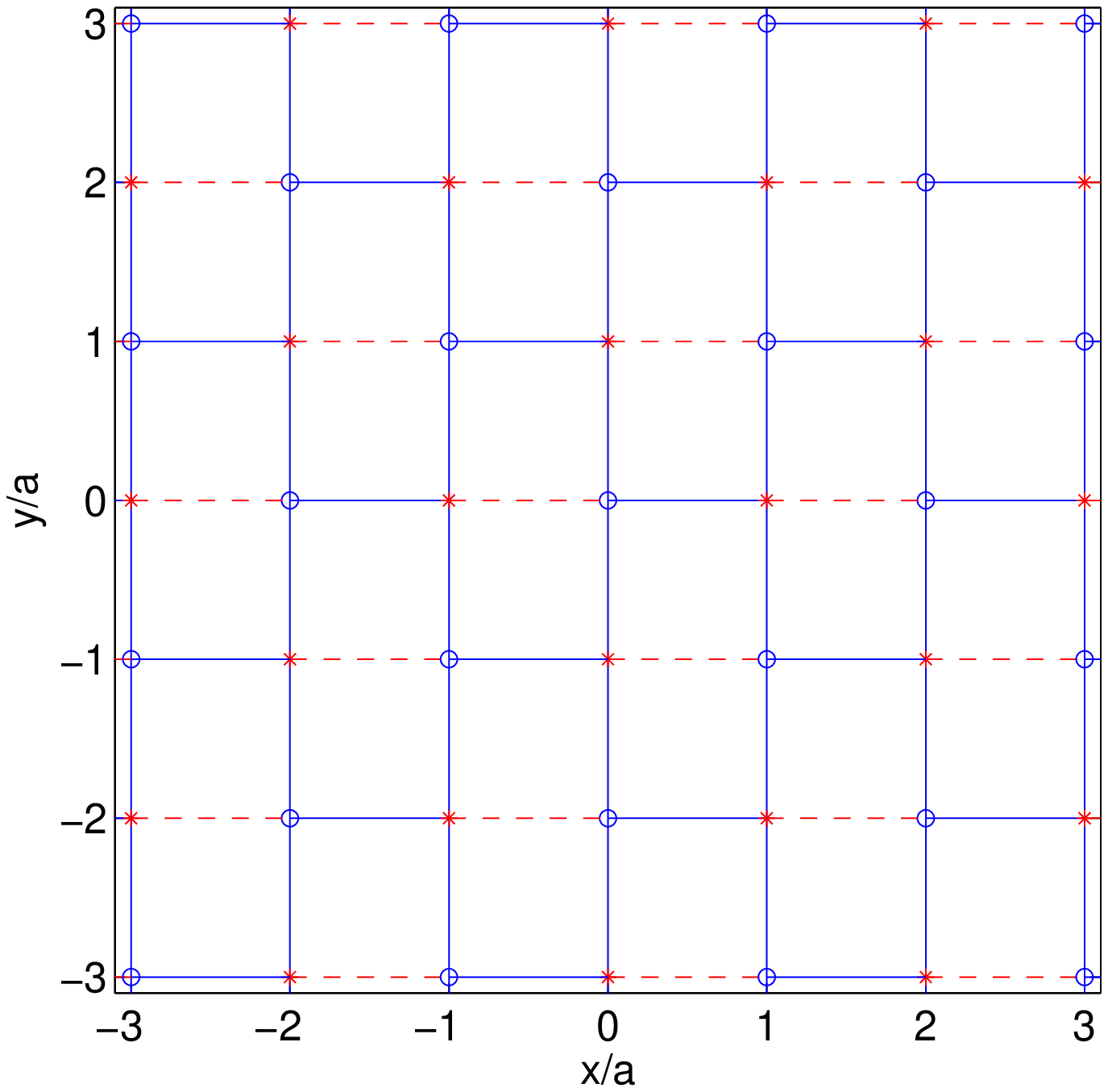}
\caption{The tight-binding model: a quantum particle has on-site energies $\pm\delta$ for the $A$ (open circles) and $B$ (asterisks) sites and can tunnel out a given lattice site to the nearest site with the tunneling rate $J_1$ (solid line bonds) and $J_2$ (dashed line bonds).}
\label{fig1}
\end{figure}

\begin{figure}
\center
\includegraphics[width=7.5cm,clip]{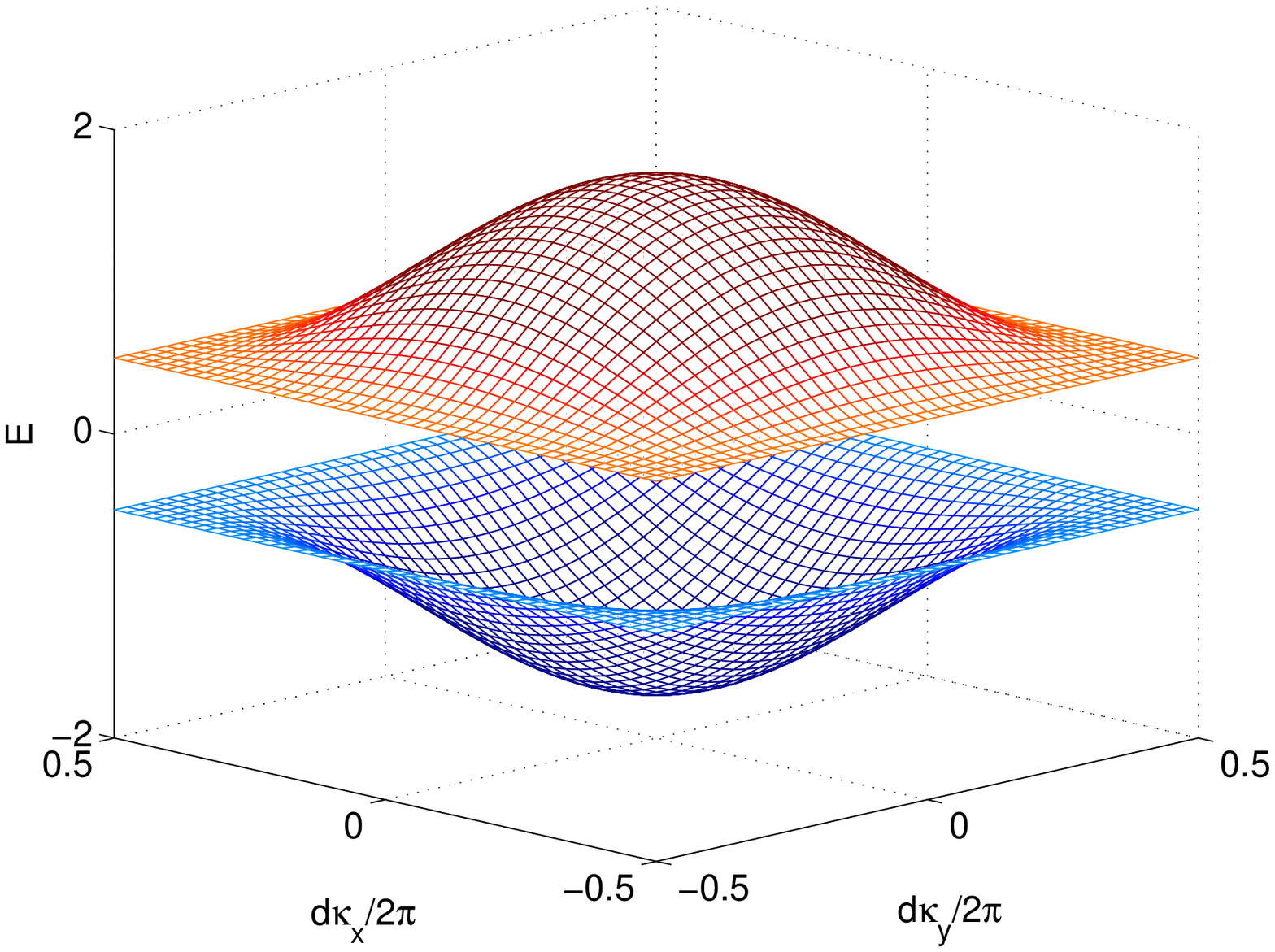}
\includegraphics[width=7.5cm,clip]{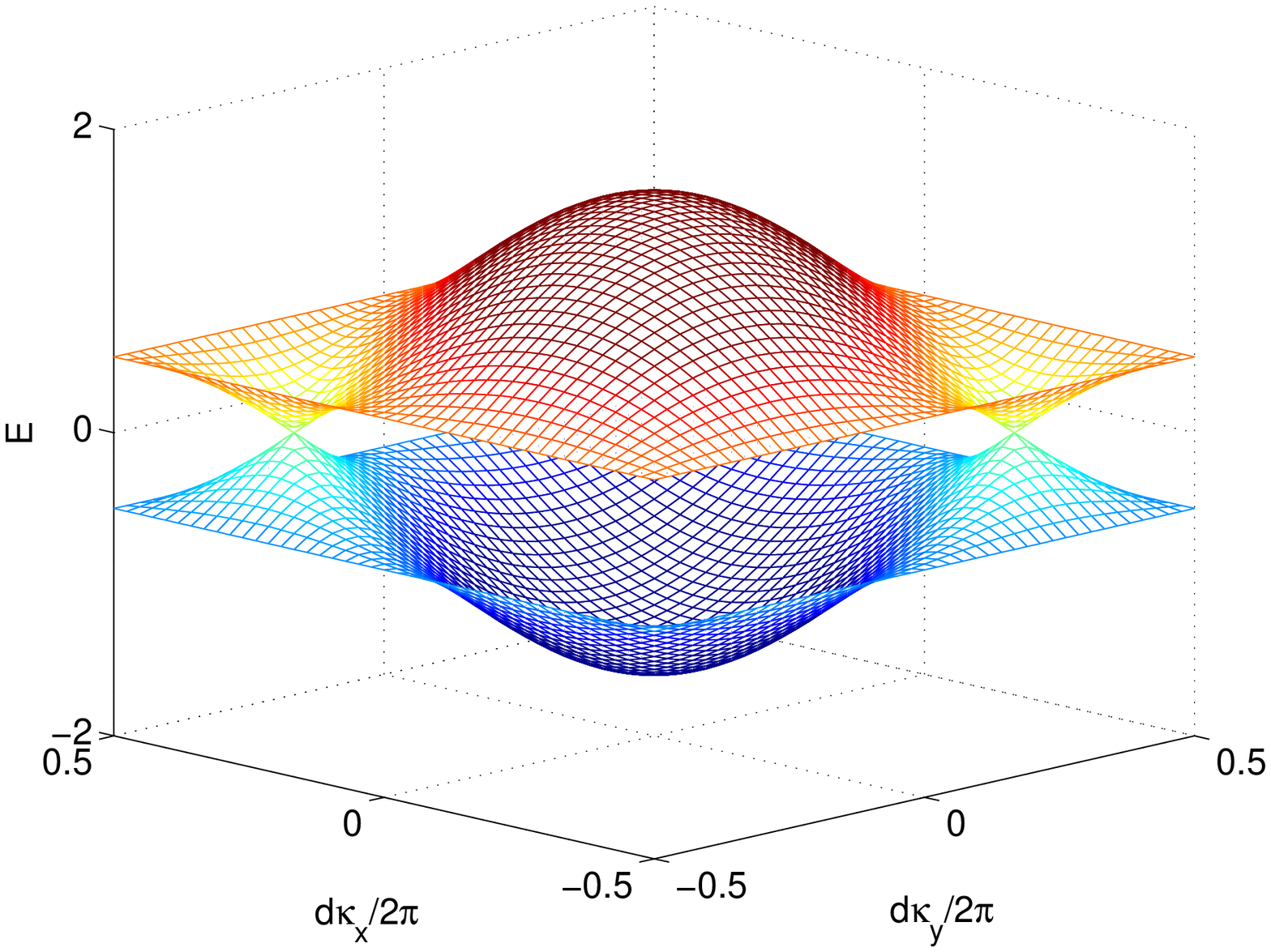}
\caption{Bloch bands of the double-periodic lattice shown in Fig.~\ref{fig1} for $J_1=J_2=0.4$ and $\delta=0.5$ (the lattice (i) in what follows), left panel , and $J_1=0.5$ and $J_2=\delta=0$ (the lattice (ii) in what follows), right panel.}
\label{fig2}
\end{figure}

As the model we shall consider a square lattice of the unit period with two different tunneling elements along the bonds and different on-site energies for the $A$ and $B$ sites, see Fig.~\ref{fig1}. Notice that the primary  axes of this lattice are rotated by 45 degrees with respect to the $xy$ axes and the new period is $a=\sqrt{2}$.  The Bloch spectrum of this lattice consists of two subbands with the dispersion relations $E=E_\pm(\kappa_x,\kappa_y)$ found as eigenvalues of the matrix
\begin{equation}
\label{b1}
H_0=\left(
\begin{array}{cc}
-\delta&-J_2-J_1(e^{-ia\kappa_x}+e^{-ia\kappa_y}+e^{-ia(\kappa_x+\kappa_y)})\\
-J_2-J_1(e^{ia\kappa_x}+e^{ia\kappa_y}+e^{ia(\kappa_x+\kappa_y)})&\delta
\end{array} \right) \;,
\end{equation}
where $\kappa_x$ and $\kappa_y$ are the Bloch wave quasimomenta. Two particular cases to be addressed in this work,  are (i) $J_1=J_2$ and $\delta\ne 0$; and (ii) $\delta=0$ and $J_2=0$.  In the former case two Bloch bands are  separated by a finite energy gap, see Fig.~\ref{fig2}(a),  while in the latter case they touch each other at the Dirac points, see Fig.~\ref{fig2}(b).  It appears that the absence or presence of Dirac's cones in the Bloch spectrum strongly  affects the Wannier-Stark spectrum and, hence, this two cases should be analyzed separately.

Let us now proceed to the tilted lattices.  As it was already mentioned in Sec.~\ref{sec1}, the fundamental difference of tilted 2D lattices as compared with tilted 1D lattices is that the Wannier-Stark spectrum and WS-states depend on the field orientation, which we shall characterized by the parameter $\beta=F_x/F_y$. Alternatively, one can characterize the field  orientation by the parameter
\begin{displaymath}
\tilde{\beta}=\tilde{F}_x/\tilde{F}_y \;,
\end{displaymath}
where $\tilde{F}_{x}$  and $\tilde{F}_{y}$ are field components in the coordinate system aligned with the primary axes of the lattice.  Then rational orientations corresponds to rational values of the parameter $\beta$  which also implies rational $\tilde{\beta}=r/q$ (here $r$ and $q$ are co-prime numbers).

\begin{figure}
\center
\includegraphics[width=10cm,clip]{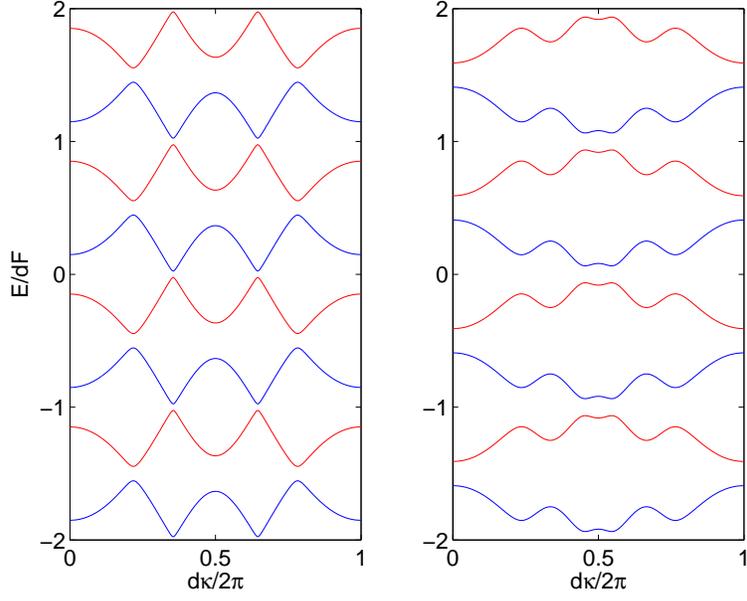}
\caption{Ladders of Wannier-Stark energy bands for the lattice (i), left panel, and the lattice (ii), right panel.  Parameters of the static field are $F=0.4$ and $F_x/F_y=1$.}
\label{fig3}
\end{figure}

\begin{figure}
\center
\includegraphics[width=7.5cm,clip]{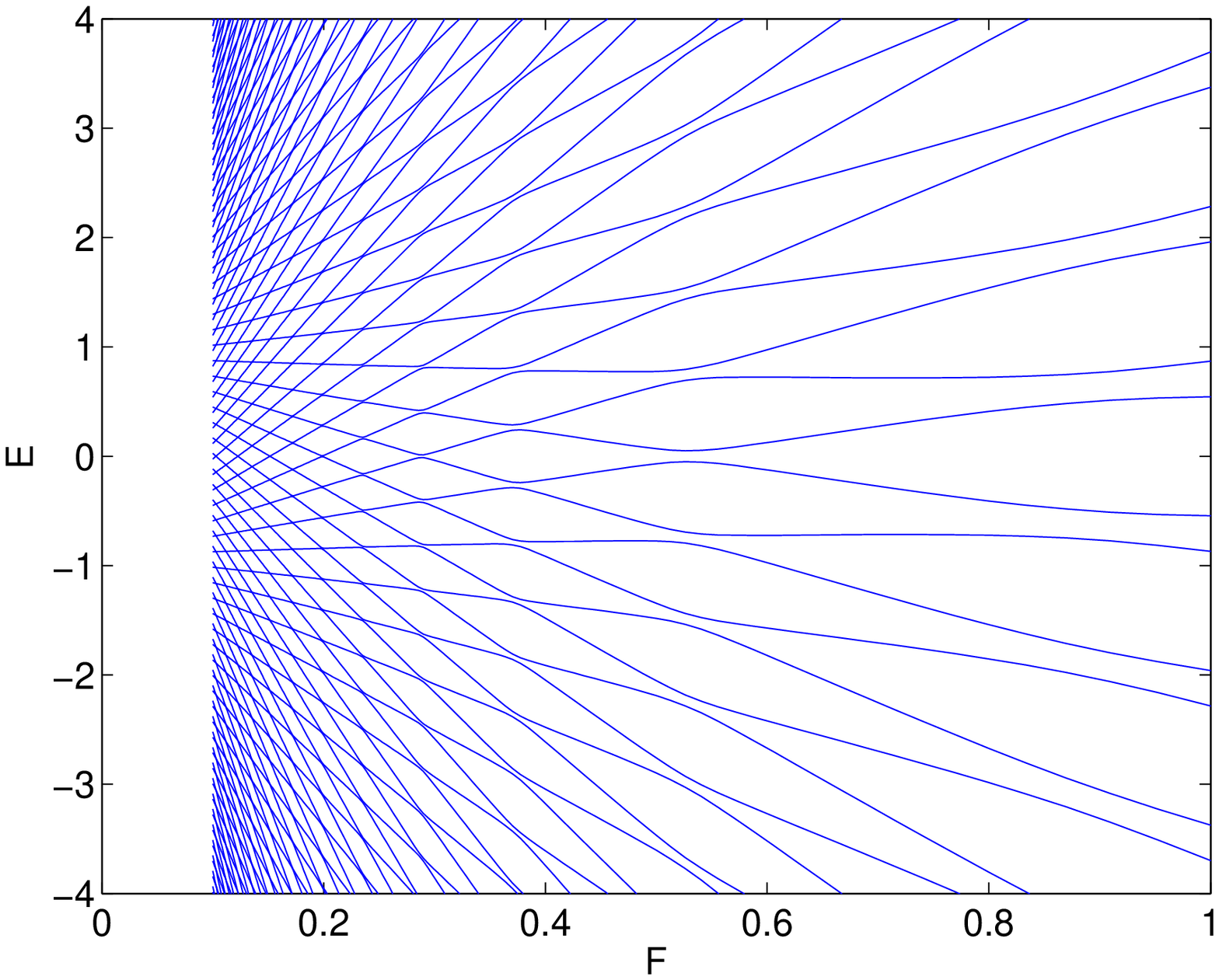}
\includegraphics[width=7.5cm,clip]{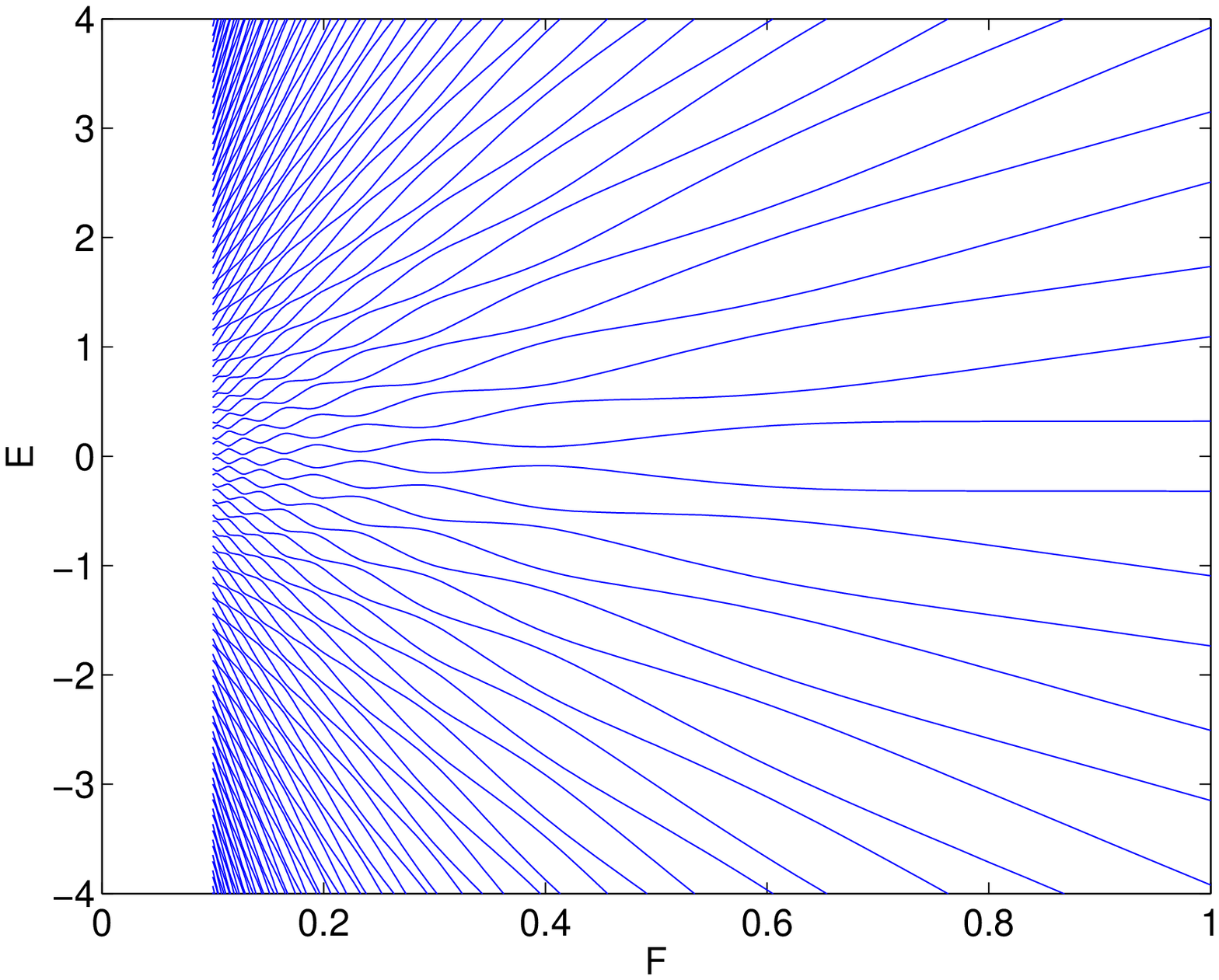}
\caption{Wannier-Stark fan for the lattice (i), left panel, and (ii), right panel. The transverse quasimomentum $\kappa=\pi/2d$ (i.e., $d\kappa/2\pi=0.25$),  $F=0.4$ and $F_x/F_y=1$. }
\label{fig4}
\end{figure}

To find the energy spectrum for rational $\tilde{\beta}=r/q$ we follow Ref.\cite{Naka93,92} and rotate the coordinate system to align the new $y$-axis with the vector ${\bf F}$, and then use the plane-wave ansatz for the wave function. This results in the following eigenvalue problem
\begin{eqnarray}
\nonumber
dF\left(l-\frac{r+q}{4}\right) \psi_l^{A}+\delta \psi_l^{A}
- J_2\psi_l^{B} - J_1\left(e^{-ird\kappa}\psi_{l-q}^{B} - e^{iqd\kappa} \psi_{l-r}^{B} - e^{id(q-r)\kappa}\psi_{l-q-r}^{B}\right)
=E\psi_l^{A} \;,\\
\label{b2}
dF\left(l+\frac{r+q}{4}\right)  \psi_l^{B}-\delta \psi_l^{B}
- J_2\psi_l^{A} - J_1\left(e^{ird\kappa}\psi_{l+q}^{A} - e^{-iqd\kappa} \psi_{l+r}^{A} - e^{-id(q-r)\kappa}\psi_{l+q+r}^{A}\right)
=E\psi_l^{A}  \;,
\end{eqnarray}
where
\begin{equation}
\nonumber
d=\frac{a}{\sqrt{N}} \;,\quad a=\sqrt{2}  \;,\quad N=r^2+q^2 \;.
\end{equation}
One obtains the WS-bands through a sweep along the plane-wave quasimomentum $\kappa$ which enters Eq.~(\ref{b2}) as a parameter. As an example Fig.~\ref{fig3} shows these bands for the parameters of Fig.~\ref{fig2} and $F=0.4$ and  $(r,q)=(1,0)$. [In the original coordinate system this orientation corresponds to $F_x/F_y=1$.] It is seen in Fig.~\ref{fig3} that WS-bands appear in pairs and are arranged into two ladders with the step $dF$. As $F$ is varied the bands change their  shape in a rather complicated way.  On the other hand, if we fix the quasimomentum $\kappa$, we obtain Wannier-Stark fan, see Fig.~\ref{fig4}, which resembles that for a 1D double-periodic lattice. In particular, it is seen in  Fig.~\ref{fig4} that energy levels of different symmetry show avoided crossing when they approach each other.  If we now  fix $F$ and vary $\kappa$ these avoided crossings become avoided crossings between energy bands. We would like to draw reader's attention that the discussed avoided crossings are much more pronounced in the case (ii) -- we come back to this point in Sec.~\ref{sec3c}.

In the next section we shall analyze WS-bands in more details, focusing on their  dependence on the static field $F$. Along with explicit form of the energy bands we shall  be also interested in the following integrated characteristic,
\begin{equation}
\label{b9}
A(F)=\frac{1}{2\pi}\int_0^{2\pi} v^2(\kappa) {\rm d}(d\kappa) \;,\quad v(\kappa)=\frac{\partial E}{\partial \kappa}
\end{equation}
where $v(\kappa)$ is the  the group velocity. It will be shown later on in Sec.\ref{sec4} that the quantity (\ref{b9}) determines the rate of ballistic spreading of a localized wave packet. For the purpose of future references Fig.~\ref{fig5} shows this quantity as the function of $1/F$.

\section{Strong {\em vs.} weak fields}
\label{sec3}

\subsection{Strong fields}
\label{sec3a}

The limit of strong fields can be solved using Rayleigh-Schr\"odinger perturbation theory with $\epsilon=1/F$ as the small parameter. As follows from Eq.~(\ref{b2}), the zero-order spectrum corresponds to flat bands arranged into two ladders.  Denoting by $E^{(r,q)}(\kappa)$ the perturbation correction, we have
\begin{equation}
\label{b4}
E_{n,\pm}(\kappa)=dF\left(n\mp\frac{r+q}{4}\right) \pm \delta \pm E^{(r,q)}(\kappa) \;.
\end{equation}
The super-index $(r,q)$ in the correction term underlines its dependence on the field orientation. It can be rigorously proved that few first terms in the perturbation series vanish and, hence,  $E^{(r,q)}(\kappa) \sim \epsilon^\nu$ where $\nu$ is number of vanishings terms.  For example, for the lattice (i) we have  $\nu=2$ if $F_x/F_y=1$, $\nu=3$ if $F_x/F_y=1/3$, etc..  Thus, in general case  the band width $\Delta$ of the WS-bands tends to zero if $F\rightarrow\infty$. Two exclusions, where the bands have a finite width in the above limit (i.e., $\nu=0$), correspond to situations where ${\bf F}$ is aligned with the $x$ or $y$ axis. It should be mentioned that the increment $\nu$ depends not only on the field orientation but also on the lattice geometry and for the lattice (ii) we have $\nu=0$ if ${\bf F}$ is parallel to the $x$ axis yet $\nu=2$ if ${\bf F}$ is parallel to the $y$ axis. It can be also shown that the first non-vanishing term of Rayleigh-Schr\"odinger perturbation theory gives the cosine dispersion relation for the WS-bands:
\begin{equation}
\label{b4b}
E^{(r,q)}(\kappa)=\Delta(F) \cos(Nd\kappa) \;, \quad \Delta(F)\sim\frac{1}{F^\nu} \;.
\end{equation}
Note that the dispersion relation $E^{(r,q)}(\kappa)$ is periodic function of $\kappa$ with the period $2\pi/Nd$ but not $2\pi/d$, as one might naively expect basing on the explicit periodicity of the coefficients in Eq.~(\ref{b2}).  Finally, we mention that the functional dependence of the quantity (\ref{b9}) on $F$ is obviously given by $A\sim (1/F)^{2\nu}$, see inset in Fig.~\ref{fig5}.

\subsection{Moderate fields}
\label{sec3b}

\begin{figure}
\center
\includegraphics[width=10cm,clip]{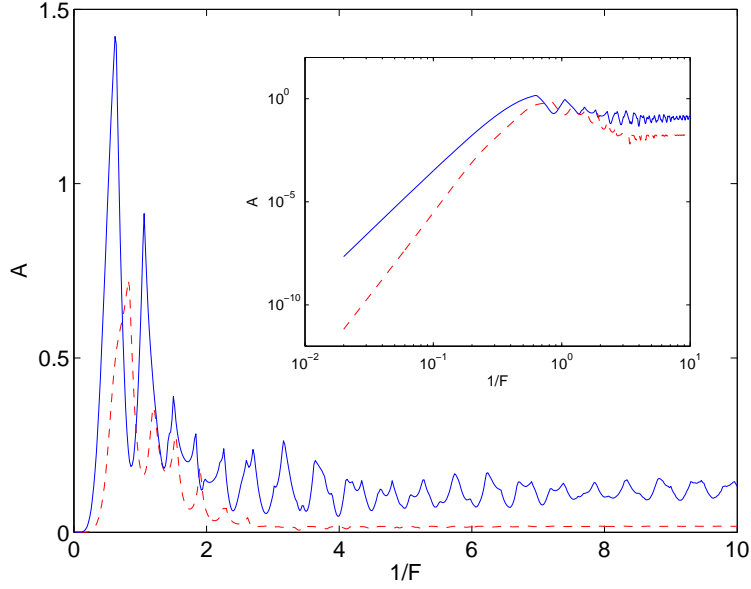}
\caption{The quantity $A$ as the function of $1/F$ for the lattice (i), dashed line, and the lattice (ii), solid line. Field orientation is $F_x/F_y=1/3$ or $(r,q)=(2,1)$. The inset shows the same functions in the logarithmic scale.}
\label{fig5}
\end{figure}

The region of applicability of the perturbation approach of Sec.~\ref{sec3a} is restricted to  small values of the parameter $\epsilon=1/F$, smaller than the position of the first peak in Fig.~\ref{fig5}. In this subsection we discuss a more sophisticated approach which allows us consider moderate $F$, where the band width $\Delta$ of WS-bands takes its maximal value. This approach is a straightforward generalization of the method used in Ref.~\cite{preprint} to analyze the Wannier-Stark spectrum of a quantum particle in one-dimensional lattices.

Similar to Eq.~(5) in Ref.~\cite{preprint} we introduce the generating functions
\begin{equation}
\label{b6a}
Y^{A,B}(\theta)=(2\pi)^{-1/2} \sum_{l=-\infty}^\infty \psi_l^{A,B}\exp(il\theta) \;.
\end{equation}
This  reduces Eq.~(\ref{b2}) to the system of two ordinary differential equations,
\begin{equation}
\label{b6}
idF\frac{{\rm d}}{{\rm d}\theta}\left(
\begin{array}{c}
Y^A\\Y^B
\end{array} \right)=G(\theta;\kappa)\left(
\begin{array}{c}
Y^A\\Y^B
\end{array} \right) \;,
\end{equation}
where
\begin{equation}
\label{b7}
G(\theta;\kappa)=\left(
\begin{array}{cc}
E+dF(r+q)/4+\delta & J_2+J_1 f(\theta;\kappa) \\
J_2+J_1f^*(\theta;\kappa) & E-dF(r+q)/4-\delta
\end{array} \right)
\end{equation}
and
\begin{equation}
\nonumber
f(\theta;\kappa)=e^{-ird\kappa}e^{-iq\theta}+e^{iqd\kappa}e^{-ir\theta} +e^{i(q-r)d\kappa}e^{-i(r+q)\theta}  \;.
\end{equation}
Equation (\ref{b6}) falls into the class of linear dynamical system and can be viewed as some effective two-level system driven by a three-chromatic field.

To solve Eq.~(\ref{b6}) one can apply a variety of methods of which the Bogoliubov-Mitropolskii averaging technique proves to be the most universal \cite{preprint}. For the lattice (ii) this techique was used in our recent paper \cite{96}, from which we borrow the final result. Restricting ourselves by one itteration, correction to the flat energy bands is given by
\begin{equation}
\label{c1}
E^{(r,q)}(\kappa)=(J_2-J_1) \sum_{n,m} {\cal J}_{m}(z_1) {\cal J}_{n}(z_2)\cos\left[\kappa d\frac{r^2+q^2}{r-q}(1+n)\right]  \;,
\end{equation}
where integer numbers $n$ and $m$ satisfy the equation $(r-q)m=-(r+q)(1+n)$ and arguments of the Bessel functions are
\begin{equation}
\nonumber
z_1=\frac{8J_1}{Fd(r-q)} \;,\quad  z_2=\frac{4(J_1+J_2)}{Fd(r+q)} \;.
\end{equation}
Notice that, if compared with Rayleigh-Schr\"odinger perturbation theory, the first order Bogoliubov-Mitropolskii correction corresponds to infinite perturbation series.

Using the fact that ${\cal J}_n(z)\sim z^n$ for $z\ll 1$ it is easy to prove that in the limit $F\rightarrow\infty$ Eq.~(\ref{c1}) recovers Eq.~(\ref{b4b}). However, unlike in Rayleigh-Schr\"odinger perturbation theory, Eq.~(\ref{c1}) captures oscillatory behavior of the band width $\Delta=\Delta(F)$, as well as the presence of higher harmonics in the dispersion relation. For example, in the above considered case $(r,q)=(2,1)$ Eq.~(\ref{c1}) takes the form
\begin{eqnarray}
\nonumber
E^{(r,q)}(\kappa)=\pm (J_1-J_2)\left[{\cal J}_0(z_1){\cal J}_1(z_2)+{\cal J}_3(z_1){\cal J}_0(z_2) \cos(5\kappa d) \right. \\
\label{c3}
\left. -{\cal J}_3(z_1){\cal J}_2(z_2)\cos(5\kappa d)- {\cal J}_6(z_1){\cal J}_1(z_2)\cos(10\kappa d)  + \ldots \right] \;.
\end{eqnarray}
Comparing Eq.~(\ref{c3}) with the exact numerical results we found that for the parameters of Fig.~\ref{fig5} it correctly approximates WS-bands  till $F\approx1$, where the quantity $A$ has the first minima.

\subsection{Weak fields}
\label{sec3c}

The opposite limit of small $F$ is more involved in the sense that the result depends on the presence of Dirac's points in the Bloch spectrum. First we shall discuss the lattice (i) where the Bloch bands are separated by the gap equal or  larger than $2\delta$.

\subsubsection{Finite gap}

A finite gap insures vanishing interband  LZ-tunneling in the limit  $F\rightarrow0$. Thus the spectrum of 2D WS-states can be found by using the adiabatic theorem, in full analogy with the weak field case for 1D lattices \cite{preprint}. Denoting by $\tilde{ {\cal E}}_\pm(\theta;\kappa)$ and ${\bf y}_\pm(\theta;\kappa)$ the eigenvalues and eigenfunctions of $2\times2$ matrix (\ref{b7}) for $E=0$, the energy spectrum of WS-states is given by the equation
\begin{equation}
\label{c6}
E_{n,\pm}(\kappa)=C_\pm(\kappa)+dF[n + c_\pm(\kappa)] \;,
\end{equation}
where
\begin{equation}
\label{c7}
C_\pm(\kappa)=\frac{1}{2\pi}\int_0^{2\pi}\tilde{ {\cal E}}_\pm(\theta;\kappa)  {\rm d}\theta \;,\quad
c_\pm(\kappa)=\frac{i}{2\pi}\int_0^{2\pi} {\bf y}_\pm^T(\theta;\kappa)\frac{{\rm d}}{{\rm d}\theta} {\bf y}_\pm(\theta;\kappa) {\rm d}\theta'
\end{equation}
are the dynamical and geometrical phases, respectively. Equation (\ref{c6}) contains the transverse quasimomentum $\kappa$ as a parameter. In fact, one can related this quasimomentum and the constants (\ref{c7}) to the Bloch spectrum $E_\pm(\kappa_x,\kappa_y)$ of the system. To do this we rotate the axis in the quasimomentum space to align the new $\kappa'_y$ axis with the vector ${\bf F}$ of the static field. Then the transverse quasimomentum $\kappa$ is associated with $\kappa'_x$ and the constant $C_\pm(\kappa)$ coinside in the limit $F\rightarrow0$ with the mean energy of the Bloch subbands along the line parallel to  ${\bf F}$:
\begin{equation}
\label{c8}
\lim_{F\rightarrow0} C_\pm(\kappa)=\frac{1}{2\pi}\int_0^{2\pi} E_\pm(\kappa,\kappa'_y)   {\rm d}(d\kappa'_y) \equiv \pm E^{(r,q)}(\kappa)  \;.
\end{equation}

\begin{figure}
\center
\includegraphics[width=10cm,clip]{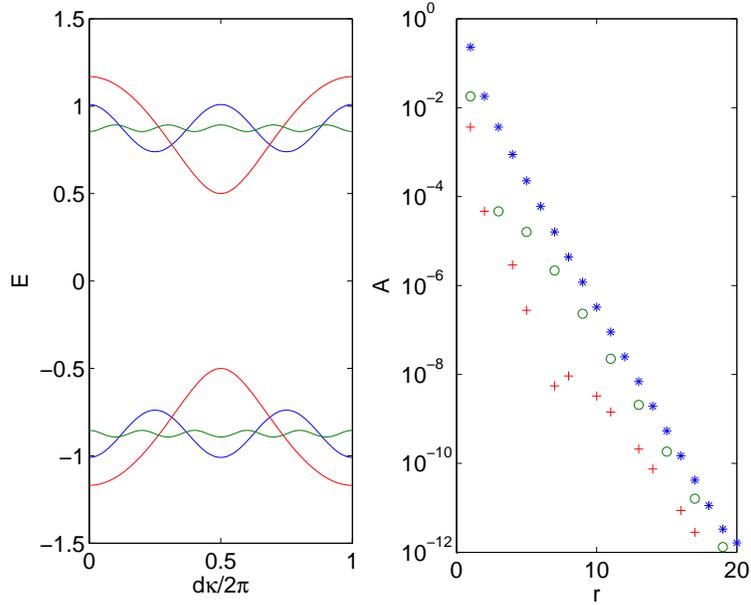}
\caption{Left panel: Dispersion relations $E^{(r,q)}(\kappa)$ in the limit $F\rightarrow0$ for the lattice (i) and $(r,q)=(1,0)$, dash-dotted line, $(r,q)=(1,1)$, dashed line, and $(r,q)=(2,1)$, solid line. Right panel: The quantity (\ref{b9}) calculated on the basis of the limiting  dispersion relation for $1\le r\le 20$ and $q=1$, asterisks, $q=2$, open circles, and $q=3$, pluses.}
\label{fig6}
\end{figure}

Let us discuss the dispersion relation (\ref{c8}) in some more details. The left panel in Fig.~\ref{fig6} shows this dispersion relation for three different orientations of the static field: $(r,q)=(1,0)$, $(r,q)=(1,1)$, and  $(r,q)=(2,1)$. It is seen that $E^{(r,q)}(\kappa)$ rapidly converges to a flat band when $r$ and $q$ are increased.  More elaborated numerical analysis shows that this convergence is exponentially fast, see riht panel inFig.~\ref{fig6}. Thus, for most of the field orientations the band width $\Delta$ takes an exponentially small value in the limit $F\rightarrow0$.


\subsubsection{Dirac's cones}
\label{sec3cc}

As seen in Fig.~\ref{fig5}, for the lattice (i) the quantity (\ref{b9}), which characterizes the width of WS-bands, takes a constant value as $F$ is decreased. Obviously, this is a manifestation  of transition to the adiabatic regime of interband LZ-tunneling. This result, however,  does not hold for the lattice (ii) where the Bloch subbands are connected at the Dirac points. In this situation we have non-negligible LZ-tunneling for arbitrary small $F$ and arbitrary field orientations \cite{Lib12}. Correspondently, the function $A=A(F)$ does not take a constant value in the limit of small $F$.

\begin{figure}
\center
\includegraphics[width=10cm,clip]{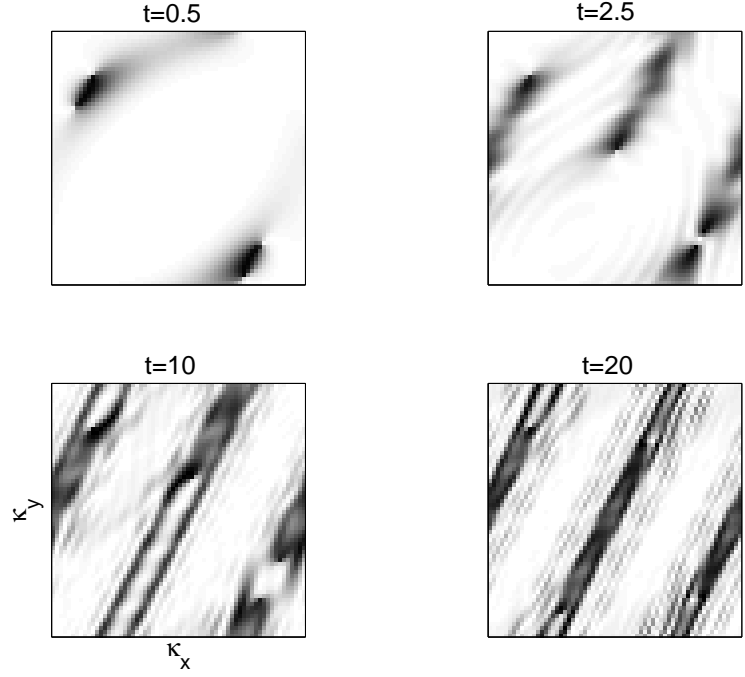}
\caption{Snapshots of $P_{+}(\kappa_x,\kappa_y;t)$ for the lattice (ii) where initially the lower Bloch subband is completely populated. Parameters are $F=0.2$, and $F_x/F_y=1/3$, the time is measured in the units of $T_J=2\pi/J_1$.}
\label{fig7}
\end{figure}

\begin{figure}
\center
\includegraphics[width=11cm,clip]{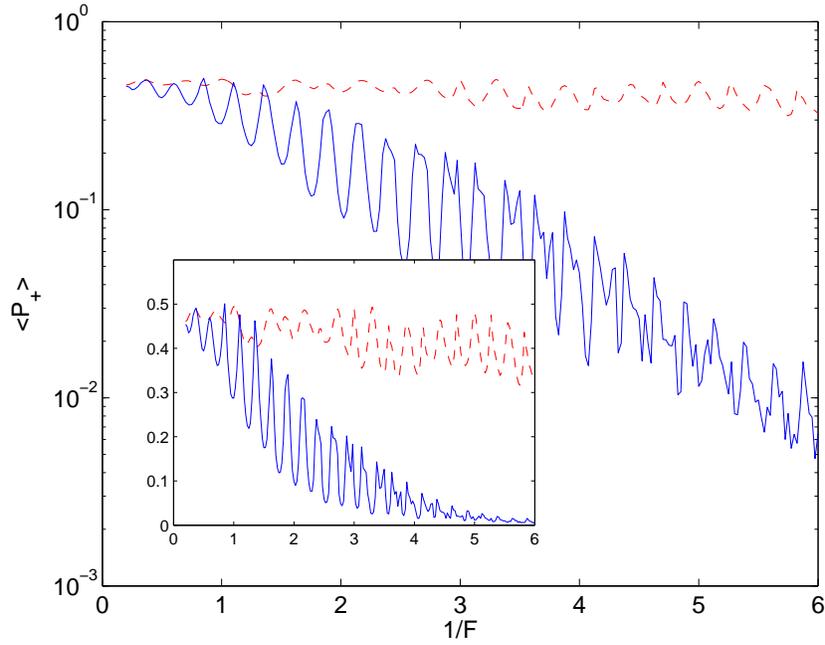}
\caption{Time-averaged population of the upper band as the function of $1/F$ for the lattice (i), solid line, and the lattice (ii), dashed line, in the linear (inset) and logarithmic (main panel) scales.}
\label{fig17}
\end{figure}

To study the interband LZ-tunneling we consider a delocalized initial condition given by a Bloch wave. For this initial condition the problem reduces to solving the Schr\"odinger equation
\begin{equation}
\label{c9}
i\frac{{\rm d}}{{\rm d}t}\left(
\begin{array}{c}
\psi^A\\\psi^B
\end{array} \right)=H(t)\left(
\begin{array}{c}
\psi^A\\\psi^B
\end{array} \right) \;,
\end{equation}
where $H(t)$ is obtained from the Hamiltonian (\ref{b1}) by substituting $\kappa_{x,y}\rightarrow \kappa_{x,y} - \tilde{F}_{x,y}t$. Notice that the solution depends on the initial quasimomenta. Thus, at any given time populations of the Bloch sub-bands are  functions of $\kappa_x$ and $\kappa_y$, i.e.,  $P_\pm(t)=P_\pm(\kappa_x,\kappa_y;t)$. Figure \ref{fig7} shows snapshots of $P_{+}(\kappa_x,\kappa_y;t)$ for the lattice (ii) where initially only the lower band was populated -- a setup corresponding to Fermi particles with the Fermi energy at $E=0$ \cite{Tarr12}. It is seen in the figure that the particles are transmitted into the upper band through the cones . In addition to Fig.~\ref{fig7}, Fig.~\ref{fig17} compares the averaged over the time total populations of the upper band for the lattice (i) and (ii). It is seen that in the former case $\langle P_+\rangle\propto \exp(-1/F)$ while in the latter case  $\langle P_+\rangle\propto const$. This explans qualitative difference between the lattice (i) and (ii) in the weak field limit.

\section{Ballistic spreading of a localized wave packet}
\label{sec4}

\begin{figure}
\center
\includegraphics[width=10cm,clip]{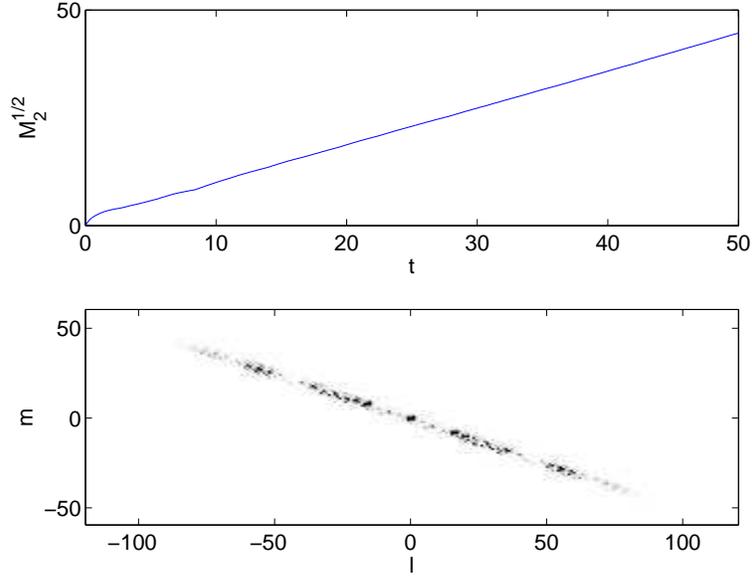}
\caption{Upper panel: Square root of the second momentum as the function of time. Lower panel: Populations of the lattice sites $|\psi_{l,m}|^2$ at $t=50T$, $T=2\pi$. Parameters are $F=0.8$, and $(r,q)=(2,1)$.}
\label{fig8}
\end{figure}

The continuous spectrum of WS-states implies ballistic spreading of a localized wave packet where, asymptotically,  the second moment of the packet grows quadratically in time:
\begin{equation}
\nonumber
M_2(t) \rightarrow B t^2 \;,\quad M_2(t)=\sum_{l,m} (l^2+m^2) |\psi_{l,m}(t)|^2 \;.
\end{equation}
For the lattice (i) this behavior is illustrated in Fig.~\ref{fig8} where initially single site at the lattice origin was populated. We mention that we shall get essentially the same result if we chose a finite-width incoherent packet as the initial condition. Approximating numerical data in Fig.~\ref{fig8}(a) by a straight line we extract coefficient $B$.

Let us now evaluate the second momentum analytically. For simplicity we shall consider the case $(r,q)=(1,1)$ where the wave packet spreads in the $x$ directions. Expending the initial wave function over the basis of WS-states we obtain,
\begin{equation}
\label{b11}
\psi_{l,m}(t)=\sum_n\int {\rm d}\kappa a_n(\kappa) \Psi_{l,m}^{(n,\kappa)} e^{-i[dFn+E(\kappa)]t} \;,
\end{equation}
where $a_{n}(\kappa)$ are the expansion coefficients and we omit plus-minus index  for two different Wannier-Stark ladders. Because WS-states are Bloch-like states in the $x$ direction the following decomposition is valid: $\Psi_{l,m}^{(n,\kappa)}=b_m^{(n)}(\kappa)e^{i\kappa l}$.
%
Next, using the identity  $l\Psi_{l,m}^{(n,\kappa)}=-i{\bf b}_m^{(n)}(\kappa){\rm d} e^{i\kappa l}/{\rm d} \kappa$ one can prove the following intermediate result:
\begin{equation}
\label{b13}
l\psi_{l,m} \approx  t \sum_n\int {\rm d}\kappa a_n(\kappa) \frac{\partial E}{\partial \kappa} \Psi_{l,m}^{(n,\kappa)} e^{-i[dFn+E(\kappa)]t}  \;.
\end{equation}
To obtain the last equation we used integration by parts and kept only the term which grows in time. With the help of Eq.~(\ref{b13}) the second moment is given by
\begin{equation}
\nonumber
M_2(t)\approx\sum_l |l\psi_{l,m}(t)|^2=t^2 \sum_n\int {\rm d}\kappa |a_n(\kappa)|^2\left(\frac{\partial E}{\partial \kappa}\right)^2 \;.
\end{equation}
Finally, using the normalization condition $\sum_n\int {\rm d}\kappa |a_n(\kappa)|^2=1$, we estimate the second moment as
\begin{equation}
\label{b15}
M_2(t) \sim A t^2 \;,
\end{equation}
where the coefficient $A$ is defined in Eq.~(\ref{b9}). Figure \ref{fig10} compares the rate of ballistic spreading $B=B(F)$, which was obtained by using the wave-packet simulations, with the quantity $A=A(F)$, which was calculated by using the dispersion relation of WS-states. An excellent agreement is noticed, in spite of a number of approximations that are used to derive Eq.~(\ref{b15}).

It is interesting to compare the wave-packet dispersion $\sigma(t)=\sqrt{M_2(t)}$ for different orientations of the vector ${\bf F}$,  see Fig.~\ref{fig10}. In the considered case of weak field $F<J$ the short-time dynamics is defined by the ballistic spreading of a localized packet for $F=0$. However, for longer time the system `begins to feel' the static field. At this stage it differentiates between rational and irrational orientations. For rational orientations the packet continues to spread in the direction orthogonal to the vector ${\bf F}$,  while for irrational orientations dispersion $\sigma(t)$ saturates at some level.  Considering the dispersion as the  function of field orientation $\theta=\arctan(F_x/F_y)$, we observe development of a fractal structure where each peak is associated with some rational $\beta=F_x/F_y$ (see Fig.~\ref{fig11}).

\section{Interband dynamics}
\label{sec5}

As mentioned in Sec.~\ref{sec1}, cold atoms in optical lattices offers a possibility for measuring populations of the Bloch subbands. In this section we discuss population dynamics and relate it to the energy spectrum of WS-states. We shall consider two different  initial conditions:  Bloch wave with zero quasimomenum (i.e., the ground state of the system) and equal occupation of all quasimomentum states belonging to the lower subband.  These initial conditions are realized in experiments with noninteracting Bose and Fermi atoms, respectively.  On the formal level the latter case implies analysis of the population dynamics for arbitrary  initial quasimomentum, followed by averaging over the Brillouin zone.

The thin solid lines in Fig.~\ref{fig18} show the total population of the upper Bloch band $P_+=P_+(t)$ for the bosonic initial condition, where two panels refer to rational orientation $F_x/F_y=1/3$ and irrational $F_x/F_y=(\sqrt{5}-1)/4\approx1/3$. It is seen that  for the rational orientation $P_+(t)$ is (almost) periodic process while it is a complex quasiperiodic process for the irrational orientation. This is explained by the fact that one-dimensional dispersion relation $E_\pm(t)=E_\pm(\kappa_x+F_x t,\kappa_y+F_y t)$   -- the quantity which implicitly enters the Schr\"odinger Eq.~(\ref{c9}) --  is periodic function of time for rational $\beta=F_x/F_y$ and quasiperodic function for irrational $\beta$. Thus for rational $\beta$ one has chance to observe periodic population dynamics similar to periodic St\"uckelberg oscillations in one-dimensional lattices.

Remarkably, for the fermionic  initial condition situation is inverted. Because of continuous spectrum of WS-states for rational $\beta$, the averaging over the Brillouin zone leads to irreversible decay of oscillations, so that population of the upper band stabilizes at some level. On the contrary, for irrational $\beta$ the spectrum of WS-states is discrete. As a consequence, the characteristic quasiperiod of $P_+(t)$ is the same for any initial quasimomentum and the averaging over the Brillouin zone single out this quasiperiod, see thick line in Fig.~\ref{fig18}(b).

\section{Conclusions}
\label{sec6}

In the work we analyzed the energy spectrum of a quantum particle in tilted double-periodic lattices with the square symmetry. It was shown that for rational orientations of a static field, $\tilde{F}_x/\tilde{F}_y=r/q$ where $r$ and $q$ are co-prime numbers, the spectrum consists of two ladders of energy bands termed the Wannier-Stark bands (WS-bands). We obtained asymptotic expressions for WS-bands in the limit of strong and weak fields and numerically analyze these bands for intermediate $F$.

As the main result we proved that WS-bands determine the rate of ballistic spreading of a localized wave-packet,  which is the quantity that can be directly measured in a laboratory experiment. Importantly, this spreading takes place only in the direction orthogonal to the vector ${\bf F}$. The underlying phenomenon behind this effect is the interband Landau-Zenner tunneling (LZ-tunneling) between Bloch subbands \cite{58}. Thus the present work can be  considered as  an analysis of interband LZ-tunneling in two-dimentional lattices -- the problem which has attracted much attention in 1D lattices \cite{Breid06,Drei09,Kling10,Plot11}.

With respect to the wave-packet spreading the results can be summarized as follows. In the strong field limit the rate of ballistic spreading $A$ obeys universal law $A\sim 1/F^\nu$ where the increment  $\nu$ is an integer number uniquely determined by the lattice geometry and field orientation  relative to the primary axes of the lattice. In the opposite weak field limit  the rate $A$ also shows universal behavior, however, only in the case where the Bloch spectrum of the system does not contain Dirac points. If it is the case, $A$ becomes independent of $F$ and depends only on the lattice geometry and field orientation. In terms of LZ-tunneling these two asymptotic regimes correspond to nearly diabatic and adiabatic interband dynamics. For intermediate $F$ the interband dynamics is neither diabatic no adiabatic and the rate $A$ is a complicated function of the field magnitude, showing a number of peaks and deeps. As argued in our recent paper \cite{96}, the oscillatory behavior of $A=A(F)$ is a 2D analogue of the phenomenon of dynamic localization in  tilted and driven 1D lattices \cite{Zhao91,Long06,Ecka09}.

To avoid confusions, we would like to stress that we considered 2D lattices with the Bloch spectrum consisting of two subbands and all above effects are absent in `simple' lattices with single Bloch band. (Here `simple' does not implies simplicity of the Wannier-Stark problem, see Refs.~\cite{Keck02,Szam07}, for example.) In the latter lattices the wave packet stay localized for any orientation of the vector ${\bf F}$, excluding the trivial case where ${\bf F}$ is aligned with one of two primary axis of the lattice. Unlike this situation, the wave packet in double-periodic 2D lattices spreads ballistically for every rational orientation $\theta=\arctan(r/q)$ with a well defined rate $A=A(F,\theta)$. As a consequence, the wave-packet dispersion  $\sigma=\sigma(t,\theta)$ becomes a fractal function of $\theta$ in course of time. This effect was earlier predicted in Ref.~ \cite{46} for strongly decaying systems where WS-states are quantum resonances. Here we found it in not-decaying systems, which should facilitate its experimental observation.

The authors acknowledge financial support of of Russian Academy of Sciences through the SB RAS integration project No.29 {\em Dynamics of atomic Bose-Einstein condensates in optical lattices} and the RFBR project No.15-02-00463 {\em Wannier-Stark states and Bloch oscillations of a quantum particle in a generic two-dimensional lattice}.


\newpage



\begin{figure}
\center
\includegraphics[width=10cm,clip]{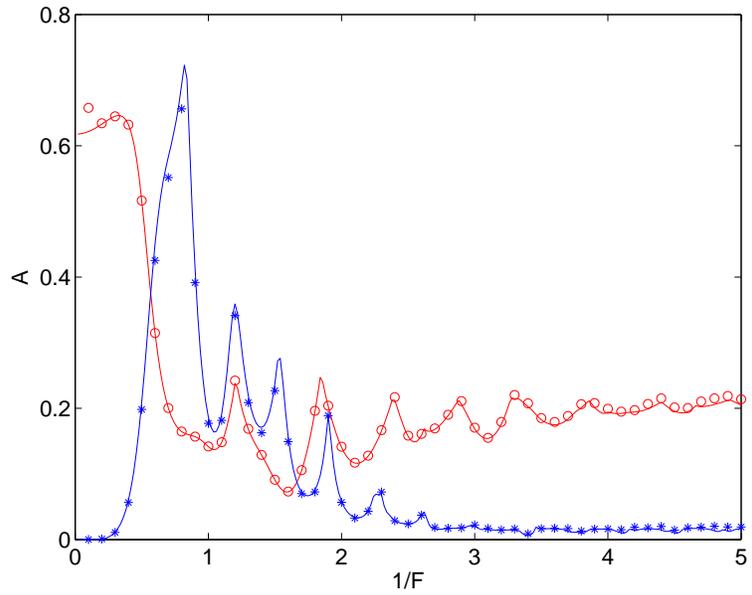}
\caption{Rates of ballistic spreading as the function of $1/F$, symbols, as compared to the analytic estimate (\ref{b9}), solid lines. The field orientation is $F_x/F_y=0$, open circles, and $F_x/F_y=1/3$, asterisks.}
\label{fig9}
\end{figure}

\begin{figure}
\center
\includegraphics[width=12cm,clip]{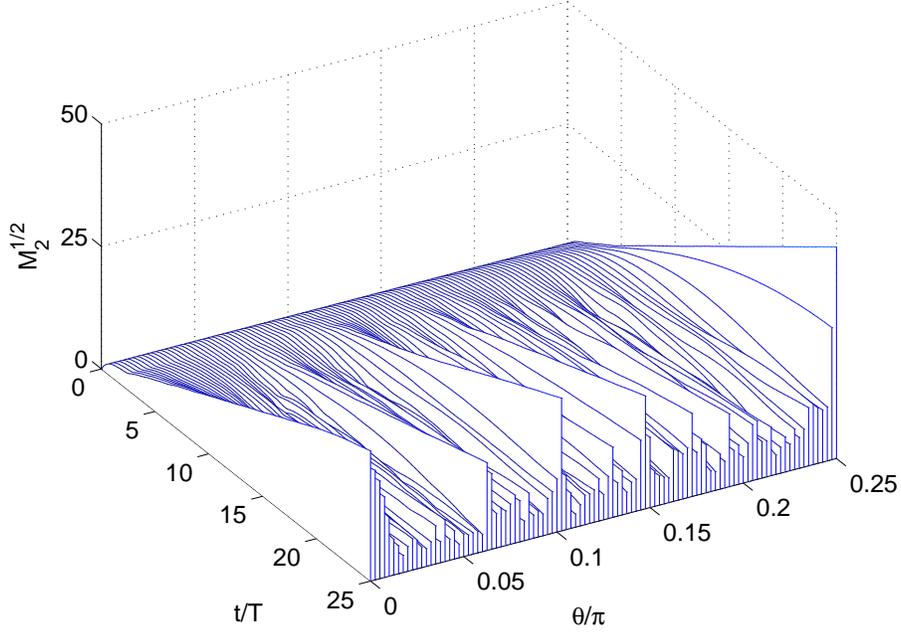}
\caption{Dispersion $\sigma(t)=\sqrt{M_2(t)}$ as the function of time for several orientations of the static field $\theta=\arctan(F_x/F_y)$. Parameters are: lattice (i), $F=0.8$.}
\label{fig10}
\end{figure}
\begin{figure}
\center
\includegraphics[width=10cm,clip]{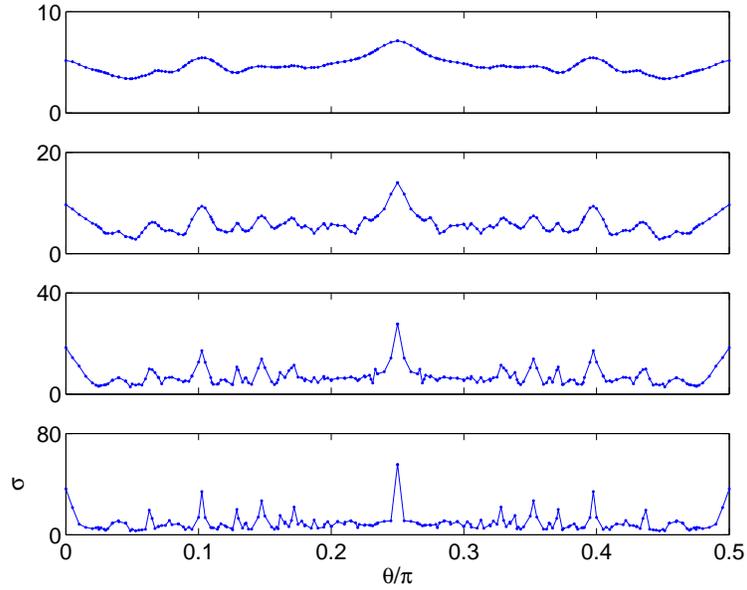}
\caption{Dispersion $\sigma(t)=\sqrt{M_2(t)}$ as the function of  $\theta$ for four different times which differ by factor 2 for the nearest panels.}
\label{fig11}
\end{figure}
\begin{figure}
\center
\includegraphics[width=11cm,clip]{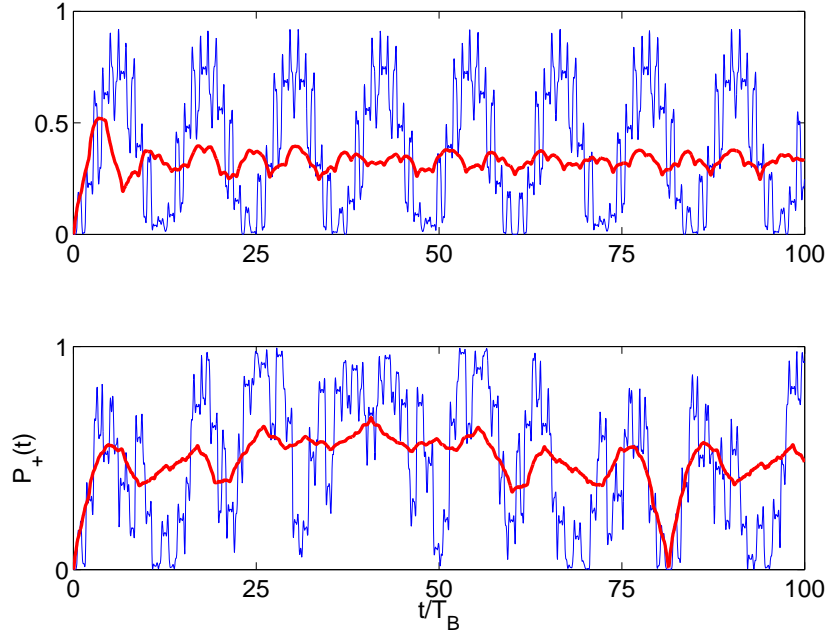}
\caption{Population of the upper band as the function of time for $\beta=1/3$, upper panel, and $\beta=(\sqrt{5}-1)/4\approx1/3$, lower panel. The solid and dashed lines refer to Bose and Fermi atoms, respectively (see text). The other parameters are: the lattice (i) and $F=0.5$.}
\label{fig18}
\end{figure}

\end{document}